# Relative Stability of Network States in Boolean Network Models of Gene Regulation in Development

**Joseph Xu Zhou[1,2]\*, Areejit Samal[1,3,4]\*, Aymeric Fouquier d'Hérouël[1,5], Nathan D. Price[1], Sui Huang[1]§**

[1] Institute for Systems Biology, Seattle, WA, USA
[2] Kavli Institute for Theoretical Physics, UC Santa Barbara, California, USA
[3] The Institute of Mathematical Sciences, Chennai, India
[4] The Abdus Salam International Centre for Theoretical Physics, Trieste, Italy
[5] Luxembourg Centre for Systems Biomedicine, Esch-sur-Alzette, Luxembourg

\*These authors contributed equally to this work
§ Corresponding author (E-mail address: shuang@systemsbiology.org)

**Abstract**

Progress in cell type reprogramming has revived the interest in Waddington's concept of the *epigenetic landscape*. Recently researchers developed the quasi-potential theory to represent the Waddington's landscape. The Quasi-potential *U(x)*, derived from interactions in the gene regulatory network (GRN) of a cell, quantifies the relative stability of network states, which determine the *effort* required for state transitions in a multi-stable dynamical system. However, quasi-potential landscapes, originally developed for continuous systems, are not suitable for discrete-valued networks which are important tools to study complex systems. In this paper, we provide a framework to quantify the landscape for discrete Boolean networks (BNs). We apply our framework to study pancreas cell differentiation where an ensemble of BN models is considered based on the structure of a minimal GRN for pancreas development. We impose biologically motivated structural constraints (corresponding to specific type of Boolean functions) and dynamical constraints (corresponding to stable attractor states) to limit the space of BN models for pancreas development. In addition, we enforce a novel functional constraint corresponding to the relative ordering of attractor states in BN models to restrict the space of BN models to the biological relevant class. We find that BNs with *canalyzing/sign-compatible* Boolean functions best capture the dynamics of pancreas cell differentiation. This framework can also determine the genes' influence on cell state transitions, and thus can facilitate the rational design of cell reprogramming protocols.

# 1. Introduction

A hallmark of multicellular organisms is the co-existence of distinct differentiated cell types with different functions and stable gene expression patterns. A less specialized cell, a stem or progenitor cell, spawns a variety of more specialized progeny cells through cell differentiation. Once differentiated, a specialized cell's gene expression pattern is relatively robust against perturbations emanating from a noisy environment. Where does this stability come from? How do gene expression patterns change as cells differentiate in response to external cues, and thereby, transition from one stable gene expression pattern to another? In principle, such questions can be answered by understanding the interactions between the genes in the underlying gene regulatory network (GRN), which constrain the changes in the gene expression patterns, producing stable and unstable steady states. The dynamical system associated with GRNs can be modelled by a system of ordinary differential equations (ODEs) where continuous variables represent the expression levels of individual genes. However, with ODEs one is quickly limited by the number of configurations of the networks due to the exponential growth of complexity with the number of genes as well as the general lack of information on the parameters that characterize the interactions between genes. A widely used alternative approach to study GRNs is Boolean networks (BNs), a framework that enables modelling of networks with hundreds of genes or analyze large statistical ensembles of networks of random structure [1,2]. Analysis of an ensemble of BNs can yield insights on the relationship between structure and dynamics of GRNs [1-3].

In 1969 Kauffman introduced BNs to study the dynamics of GRNs [1]. Since then BNs have been used to model a wide range of biological phenomena such as cell cycle, cellular differentiation and evolution of GRNs [4–17]. Specifically, BNs have been extensively used to study developmental processes. Villani *et al* [13] have developed a BN framework for cell differentiation. Krumsiek *et al* [14] have developed a BN model to recapitulate hematopoiesis. Chang *et al* [15] employed a BN model to explain human embryonic stem cell differentiation and the generation of induced pluripotent stem cells (iPSCs). Klipp *et al* [16] used a BN model to study the influence of gene regulation, methylation and histone modifications on cell differentiation. Alvarez-Buylla *et al* [6,17] used BNs to explain cell differentiation and developmental ordering in the floral organ of Arabidopsis. An important limitation of these reconstructed BN models [4-17] for different biological processes is their specification of one defined set of Boolean functions for genes in the network out of a multitude of possible choices [18] that can reproduce the biologically relevant cell states as network attractors, and the reason for the chosen set of Boolean functions often remains elusive. Also experimental observations in cell differentiation systems usually are consistent with a large number of possible Boolean functions rather than suggesting a single well-defined set, giving rise to a set of possible BNs that can describe the observed gene expression patterns of the attractors [18]. Thus, one always wonders whether the reported results would still hold for other choices of functions and how *structurally robust* the predicted dynamics is for the observed attractor states.

A more stringent requirement on a model capturing the development of multicelluar organisms is the following constraint. In addition to recapitulating the multiple observed attractors of the network, the model of the developmental GRN should also reproduce the experimentally observed *relative stabilities of attractors*, i.e, the model has to relate the different attractor states to each other based on their *relative stabilities*. By that we mean the *relative ease* for transitioning from one attractor state (*A*) to another state (*B*) which epitomizes the developmental process. More formally, in a stochastic system, the relative ease of transitioning from state *A* to state *B* would be given by the probability $P(A \rightarrow B)$ for transition from *A* to *B* (given random flutuations in gene expression). Note that such transition probabilties are typically asymmetric (i.e., $P(A \rightarrow B) \neq P(B \rightarrow A)$) – a property that ultimately accounts for the directionality (irreversibility) of development.

Interactions between genes collectively produce the developmental ordering of different cell types which is robust and repeatable during embryogenesis. Therefore, once the multiple attractors of the dynamical system are determined, it is necessary to evaluate their relative stabilities in order to derive a consistent relative ordering for all attractors in a developmental process (if one exists). Recently, some of us have derived a framework to calculate the relative stabilities of cell attractors in continuous ODE-based GRN models using least action principles [19]. However, ODE-based GRN models are not well-suited to model large networks, let alone ensembles of networks, for which BNs are commonly used [1-3].

In this paper, we present a mathematical framework for calculating the relative stabilities of cell attractors and transitions, and hence deriving the notion of a landscape in BN models of development. We use a minimal GRN for pancreas development as an example to demonstrate the utility of our method. Imposing the observed relative ordering of attractors as a novel phenotypic constraint affords evaluation of ensembles of BNs (with a given network structure but different sets of Boolean functions) that are compatible with multiple observed attractors of the GRN. Our method can be used to reconstruct simple BN models for developmental processes from available information on GRN architecture and relative stability of attractor states, and thus, can predict the efforts associated with particular state transitions of interest which in turn can facilitate the rational protocol design for cell reprogramming in regenerative medicine.

## 2. Modelling Framework

### 2.1 Boolean network (BN) model

BN model for a GRN is specified by its set of nodes, directed edges and Boolean functions. In a BN, the nodes represent genes while the edges represent interactions among genes in the network. Any gene $i$ in a BN at a given time can be in one of two expression states: *on* if its state $x_i = 1$ and *off* if its state $x_i = 0$. For a $m$-gene BN, the state vector $\mathbf{X}^t = (x_1(t), x_2(t), \ldots, x_m(t))$ gives the expression of all genes at discrete time $t$ in the network. For each gene $i$ in a BN, a Boolean function $F_i$ determines the output value $x_i$ at time $t + 1$ given the state of its input genes at time $t$. Thus, the gene expression state of a BN at any time step is governed by the recursive equation:

$$\mathbf{X}^{t+1} = \mathbf{F}(\mathbf{X}^t) \qquad (1)$$

where $\mathbf{X}^t$ is a $m$-dimensional binary vector (0 or 1) that gives the expression of all genes at time step $t$. $\mathbf{F}$ encapsulates both the network topology and Boolean functions at all nodes, and thus, contains the information determining the dynamics of the BN.

For a $m$-gene BN, there are $2^m$ possible states. A sequence of states $\mathbf{X}^0, \ldots, \mathbf{X}^t, \mathbf{X}^{t+1}, \ldots$ forms a trajectory in the state space. Trajectories converge in a deterministic (noise-free) system. Since the state space is finite, the trajectories eventually coverge either to a single state (point attractor) or a cycle of states (cyclic attractor). In the extreme case, a cyclic attractor encompasses all or almost all possible network states, and given the large number of states $2^m$, such behavior will appear *chaotic*. For any given attractor, its associated basin of attraction is the set of initial states that will converge to that attractor. Attractors of a BN are charaterized by the size (and shape) of their associated basin of attraction. The network topology (i.e., the set of nodes and edges) and the Boolean functions at each node fully determine the *attractor structure* – which consists of attractors, trajectories and basins of attraction. The attractor structure can be determined by explicitly evaluating all state transitions $\mathbf{X}^1 = \mathbf{F}(\mathbf{X}^0)$ for all $2^m$ possible initial states $\mathbf{X}^0$. An example of a 4-gene GRN with Boolean functions and resulting attractor structure is shown in Fig. 1.

**2.2 Canalyzing, Nested canalyzing and Sign-compatible Boolean functions**

Information on the topology of a GRN is increasingly available from direct experimental determination of gene regulatory mechanism or such information can be inferred from experimental data [20–23]. However, the Boolean function that controls the expression state of each gene based on the state of its input genes is more difficult to determine than network topology. In practice, a vast number of Boolean functions are plausible for a given network topology based on available experimental data [18]. Thus, it is desirable to apply biologically motivated constraints to limit the number of possible Boolean functions. We propose to use *canalyzing, nested canalyzing* and *unate* Boolean functions to charaterize the gene interactions and limit the set of possible Boolean functions for biological systems.

Kauffman proposed the concept of *canalyzing* Boolean functions (CFs) [2]. A Boolean function $F(x_1, ..., x_n)$ is CF if there exists an input $x_i$ with canalyzing Boolean value $a$ such that $F(x_1, ... x_i = a, x_n) = b$ is constant. Thus, the input value $x_i = a$ determines the output value of the Boolean function $F(x_1, ..., x_n)$ regardless of the state of other inputs. The output value $b$ is called the canalyzed value.

Later Kauffman extended this concept to *nested canalyzing* Boolean functions (NCFs) which are a subset of CFs [24]. A Boolean function $F(x_1, ..., x_n)$ is NCF if there exists a permutation of inputs $(\sigma_1, ..., \sigma_n)$ such that it satisfies the following conditions: $F(x_{\sigma_1} = a_{\sigma_1}, ..., x_{\sigma_n}) = b_{\sigma_1}$; $F(x_{\sigma_1} \neq a_{\sigma_1}, x_{\sigma_2} = a_{\sigma_2} ..., x_{\sigma_n}) = b_{\sigma_2}$; .........; $F(x_{\sigma_1} \neq a_{\sigma_1}, x_{\sigma_2} \neq a_{\sigma_2} ..., x_{\sigma_{n-1}} = a_{\sigma_{n-1}}, x_{\sigma_n}) = b_{\sigma_{n-1}}$; $F(x_{\sigma_1} \neq a_{\sigma_1}, x_{\sigma_2} \neq a_{\sigma_2} ..., x_{\sigma_{n-1}} \neq a_{\sigma_{n-1}}, x_{\sigma_n} = a_{\sigma_n}) = b_{\sigma_n}$ where $b_{\sigma_i}$ is the canalyzed Boolean value of input $x_{\sigma_i}$. The presence of CFs in BNs results in lesser number of effective upstream regulators for genes than it may appear in the network topology [24]. Furthermore, it has been shown that CFs improve the robustness of the attractor states in BNs to both mutations (in the form of changes in architecture, such as rewiring of edges or deletion of nodes) as well as perturbations (in the form of instant *bit flips* of the node states), and thus, CFs push the BN dynamics towards an *ordered* regime [24].

Another class of Boolean functions that influence the network dynamics comprises the *unate* functions. An unate function is a type of Boolean function which has monotonic properties. Since in biology a gene interaction is often characterized as either activating or inhibiting, *unate* functions can be used to capture the monotonic properties of the Boolean functions. A Boolean function $F(x_1, x_2, ..., x_n)$ is said to be *positive unate* in input $x_i$ if $\forall x_j \in \{0,1\}$ with $j \neq i$:

$$F(..., x_{i-1}, 1, x_{i+1}, ...) \geq F(..., x_{i-1}, 0, x_{i+1}, ...) \tag{2}$$

and *negative unate* in input $x_i$ if $\forall x_j \in \{0,1\}$ with $j \neq i$:

$$F(..., x_{i-1}, 1, x_{i+1}, ...) \leq F(..., x_{i-1}, 0, x_{i+1}, ...) \tag{3}$$

*Positive unateness* in input $x_i$ means that when the input gene $x_i$ is *on,* the output of the Boolean function is greater than or equal to that when the input gene $x_i$ is *off. Negative unateness* in input $x_i$ means that when the input gene $x_i$ is *on,* the output of the Boolean function is less than or equal to that when the input gene $x_i$ is *off.* Note that it has been shown earlier that NCFs are equivalent to unate cascade functions [25].

Although the Boolean function controlling the expression state of an output gene based on the state of its input genes is difficult to determine, in many cases, the nature of gene interactions (activation or inhibition) between the input gene(s) and the output gene is known from experimental data. Such information on the nature of gene interactions can be used to constrain the set of Boolean functions. Specifically, NCFs can be limited to a subset of *sign-compatible* functions (SGNs) based on such information on the nature of gene interactions. A NCF

$F(x_1, x_2, \ldots, x_n)$ is said to be a SGN if every input $x_i$ satisfies either positive unateness or negative unateness based on the known nature of interaction between input gene and output gene. If the output function for gene $x_j$ is a SGN then its input $x_i$ will satisfy positive unateness if $x_i$ is known to activate $x_j$ or its input $x_i$ will satisfy negative unateness if $x_i$ is known to inhibit $x_j$.

In summary, a hierachy exists among the different types of Boolean functions considered here. CFs are a subset of all possible Boolean functions, NCFs are a subset of CFs and SGNs are a subset of NCFs. We show later that the *sign-compatiblity* plays an important role in constraining the set of possible BNs compatible with multiple observed attractors and in determining the relative stabilities of cell attractors.

## 2.3 Transition matrix and BN dynamics

Spontaneous transitions between attractors that underlie the *epigenetic landscape* or the quasi-potential landscape requires probabilistic (noise-driven) dynamics. A discrete Markov model can be used to describe the BN dynamics. Let $p_i$ give the probability for the occupation of a state $s_i$, and $T_{ij}$ give the transition probability from state $s_j$ to state $s_i$. In a BN with $m$ genes, there are $2^m$ possible states, and the occupation probability of different possible states at time $t$ can be represented by the probability distribution $\mathbf{p}^t = (p_1^t, \ldots, p_k^t)^\mathrm{T}, (k = 2^m)$. The evolution of the discrete-time Markov model associated with the BN of $m$ genes is governed by:

$$\mathbf{p}^{t+1} = \mathbf{T}\,\mathbf{p}^t = \begin{bmatrix} T_{11} & \cdots & T_{1k} \\ \vdots & \ddots & \vdots \\ T_{k1} & \cdots & T_{kk} \end{bmatrix} \begin{pmatrix} p_1 \\ \vdots \\ p_k \end{pmatrix} \tag{4}$$

The steady state distribution is then given by eigenvector $\mathbf{p}^*$ of Markov transition matrix $\mathbf{T}$ corresponding to the eigenvalue $\lambda = 1$. Thus, BN dynamics can be characterized by the eigenvectors of Markov matrix $\mathbf{T}$. If $\mathbf{p}^*$ has only one non-zero element, then the network has a point attractor. If $\mathbf{p}^*$ has $\omega$ non-zero elements with $1 < \omega < 2^m$, then the network has cyclic attractors. If $\mathbf{p}^*$ has $2^m$ non-zero elements, it means that every state can transition to every other state, and the network dynamics is ergodic.

## 2.4 Mapping BN to a Markov model

We describe here the procedure to map a BN model to a Markov model with a simple example. Consider the toggle switch between two genes ($x$ and $y$) with both cross-inhibitory interactions and self-activation as shown in Fig. 2. The BN model of the 2-gene toggle switch can have

$2^2 = 4$ possible states: $s_1 = (0,0)$, $s_2 = (0,1)$, $s_3 = (1,0)$, and $s_4 = (1,1)$. The Boolean functions for the 2 genes in the BN model of the toggle switch are given by:

$$x^{t+1} = x^t \text{ OR NOT}(x^t \text{ OR } y^t)$$
$$y^{t+1} = y^t \text{ OR NOT}(x^t \text{ OR } y^t) \tag{5}$$

For the BN model of the toggle switch, the occupation probability of the 4 possible states is represented by a distribution vector $\mathbf{p} = (p_1, p_2, p_3, p_4)$. Based on Boolean functions in Eq. 5, the Markov transition matrix for the BN model of the toggle switch is:

$$\mathbf{T} = \begin{pmatrix} 0 & 0 & 0 & 0 \\ 0 & 1 & 0 & 0 \\ 0 & 0 & 1 & 0 \\ 1 & 0 & 0 & 1 \end{pmatrix} \tag{6}$$

where the element $T_{ij}$ gives the transition probability from state $s_j$ to state $s_i$. In the case of the toggle switch, the Markov matrix $\mathbf{T}$ has three eigenvectors with eigenvalue $\lambda = 1$. The corresponding steady states are represented by the eigenvectors $\mathbf{p}^* \in \{(0,1,0,0), (0,0,1,0), (0,0,0,1)\}$ and the three corresponding point attractors are $s_2$, $s_3$, and $s_4$.

## 2.5 Relative stability of states in BN model with noise

In a deterministic BN model, there cannot be any transitions between two states in different basins of attraction, and thus the relative stabilities of attractors have no meaning for such a Markov model. However, introduction of noise into deterministic BNs will render the possibility of transitions between two states in different basins of attraction. Thus, the notion of relative stability of the attractors can be defined in BNs with noise where there is possibility of transitions between attractors. Also, the Markov transition matrix for BNs with noise is ergodic because there is a non-zero probability of transition between any two states of the network.

In a BN with $m$ genes, any state $s_i$ in the Markov model can be represented by a $m$-bit binary vector. Let $\eta$ be the probability for randomly flipping due to noise one bit in the $m$-bit binary vector corresponding to any state $s_i$. A perturbation matrix $\mathbf{P}$ can then be constructed using the Hamming distance $d_{ij} = \|s_i - s_j\|_H$ between two states $s_i$ and $s_j$ as follows [26]:

$$P_{ij} = \begin{cases} C^m_{d_{ij}} \eta^{d_{ij}} (1-\eta)^{m-d_{ij}} & (i \neq j) \\ 0 & (i = j) \end{cases} \tag{7}$$

where $C_{d_{ij}}^n$ are binomial coefficients, which guarantee that each column of matrix **P** sums up to $1 - (1 - \eta)^m$. The BN model with noise is then constructed by adding the perturbation matrix **P** to the Markov matrix **T**:

$$\widetilde{\mathbf{T}} = (1 - \eta)^m \mathbf{T} + \mathbf{P} \tag{8}$$

The equation for the dynamics of the BN model with noise is then:

$$\mathbf{p}^{t+1} = \widetilde{\mathbf{T}} \mathbf{p}^t \tag{9}$$

Note that the Markov matrix **T** for the deterministic BN model is such that each column of **T** adds up to 1. Similarly, in Eq. 8, a normalizing coefficient is used to ensure that $\widetilde{\mathbf{T}}$ also satisfies the conservation principle, i.e., each column of $\widetilde{\mathbf{T}}$ adds up to 1. Since the Markov matrix $\widetilde{\mathbf{T}}$ is ergodic for $\eta > 0$, the BN model with noise will have one and only one steady state probability distribution $\mathbf{p}^*$ [27]. Equations 7 and 8 can be used to construct a BN model with noise which will always converge to a steady state probability distribution $\mathbf{p}^*$ that can be directly calculated from Eq. 9. With the introduction of noise, the original dynamical system corresponding to the deterministic BN relaxes into an ergodic dynamical system, and we can quantify the relative stabilities of distinct attractors for such a system via two measures: steady states probability distribution $\mathbf{p}^*$ and mean first passage time (MFPT).

Since any GRN is an open (non-conserved) system far from thermal equilibrium, the steady state probability distribution $\mathbf{p}^*$ is usually a non-equilibrium steady state with presence of circular (or non-gradient) driving forces [19]. In cases where circular driving forces are negligible compared to the gradient forces which drive a state towards point attractors, the quasi-potential function $U \propto -\sum_i \ln p_i^*$ can be used to measure the relative stability. If $U$ at one attractor is higher than other attractors, the attractor is less stable than the others, and vice versa. By contrast, if the circular driving forces are comparable to the gradient forces, $U \propto -\sum_i \ln p_i^*$ will not be an accurate measure for relative stability. When circular driving forces are large, the dynamical system has no consistent relative ordering of attractors that can be defined by $U$ because the gradient $-\frac{\partial U}{\partial x}$ is no longer the main hindrance force against the cell state transition [19]. Instead, one can use a pair-wise relationship defined based on mean first passage times (MFPTs) to measure the relative stability of cell states.

In an ergodic dynamical system, it is possible to reach every possible state from any arbitrary state. A transition barrier that epitomizes the ease of transition from one attractor to another attractor offers a notion of relative stability. Relative stability can be measured via MFPT which evaluates the average number of time steps $M_{ij}$ that are necessary for the system to transition from state $s_i$ to state $s_j$ [28].

## 2.6 Relative transitive ordering in a developmental GRN

Cell differentiation is a largely (spontaneously) irreversible process where the zygote differentiates robustly into other cell types in a pre-determined order. This programmed behavior that allows the zygote to robustly differentiate into other cell types in a well-defined sequence in the presence of molecular noise requires that the GRN orchestrates a consistent relative ordering of different attractors and intermediate transient states. By consistent relative ordering of different attractors, we mean that in the directed graph with each node as attractor state and edges representing spontaneous attractor transitions, there is little or ideally no circular (non-transitive) structure. Based on the relative stabilities of cell states measured via MFPT, we propose below a mathematical formula to define a consistent relative ordering of all attractors in the GRN.

Suppose we have a relative stability matrix **M** which gives the transition barrier between any two states based on MFPT, then we can define a relative ordering as follows. First, we define the net transition rate $D_{ij}$ between two attractors, $i$ and $j$, as follows:

$$D_{ij} = \frac{1}{M_{ij}} - \frac{1}{M_{ji}} \tag{10}$$

Note that $D_{ij} > 0$ implies that attractor $i$ is more stable than attractor $j$. This relationship can be extended to $n$ different attractors of the network. From the structure of the skew-symmetric matrix **D**, we determine whether the set of $n$ attractors is ordered with respect to the relative stability of attractors $i$ and $j$, or whether cycles are present as follows. If there exists an even permutation $\boldsymbol{\pi} = (\pi_1, \ldots, \pi_n): \{1, \ldots, n\} \to \{1, \ldots, n\}$ with $\text{sgn}(\boldsymbol{\pi}) = 1$ and $\forall i \neq j: \pi_i \neq \pi_j$, such that $\forall \pi_i < \pi_j: D_{\pi_i \pi_j} \geq 0$, then $\boldsymbol{\pi}$ defines a consistent relative ordering. Note that when a set of states has consistent ordering then there are no cycles in terms of relative stability between the states.

## 3. Results and Discussion

### 3.1 BN model for pancreas differentiation with restriction of Boolean functions

In this work, we apply the concept of relative stability of attractor states to study cell differentiation in BN models of human pancreas development.

The pancreas is an exocrine gland that secretes various digestion enzymes into the intestinal lumen. It is also an important endocrine organ: $\beta$ cells of the pancreas islets secrete insulin into the circulatory system that regulates blood glucose levels and metabolism as shown in Fig. 3A. As deficiency or malfunction of $\beta$ cells leads to diabetes, a major focus of research has been on cell regeneration or reprogramming to produce $\beta$ cells. This line of research has led to the characterization of cell lineage determining factors in the pancreas development as well as reprogramming experiments to β cells. *Pdx1* is the first gene in embryonic development that marks the onset of the pancreas cell lineage. The cells with high *Ptf1a* expression are fated to the exocrine pancreas. A few cells that display temporarily high *Ngn3* expression will form the endocrine cell lineages. The expression of *Pax4* in some cells further specifies the $β, δ$ lineages while *Arx* specifies the $α$,PP lineages (Fig. 3B).

Based on experimental literature, we reconstruct a minimal GRN for pancreas development shown in Fig. 3C that determines three cell types: the exocrine cells, the $β/δ$ progenitor and the $α$/PP progenitor. The set of interactions between the five genes in the minimal GRN for pancreas development shown in Fig. 3C were deduced from the literature on ChIP-chip or knockout or overexpression experiments [29–34]. Supplementary Table S1 lists literature evidence in support of interactions between genes contained in the minimal GRN for pancreas development. Note that in Fig. 3C the interactions shown with continuous lines have known signs (activation or inhibition) based on experimental data while interactions shown with dashed lines have unknown signs. In principle, interactions with unknown signs can either mediate activation or inhibition.

We decided to build a BN model for pancreas development based on the structure of the minimal GRN of 5 genes along with known nature of interactions as shown in Fig. 3C. Since, the number of possible Boolean functions for a node with $k$ inputs is $2^{2^k}$, the total number of all possible combinations of Boolean functions for the minimal GRN of 5 genes shown in Fig. 3C is $2^{2^1} \times 2^{2^3} \times 2^{2^3} . 2^{2^3} \times 2^{2^3} \approx 1.7 \times 10^{10}$ (Table 1), and this number is too large for any practical computational purpose. Thus, we decided to impose (structural) constraints on the type of functions to limit the set of possible Boolean functions in the BN model of the pancreas development.

As a first constraint, we explicitly adopt Boolean functions of the following type: CFs, NCFs and SGNs (See Section 2.2). Such functions may render pancreas development to be an ordered and directional process. Since, the number of possible CFs for a node with $k=1$ input is 4 and with $k=3$ inputs is 120, the total number of all possible combinations of CFs for the minimal GRN of 5 genes is $4 \times 120^4 \approx 8.3 \times 10^8$ (Table 1). Similarly, the number of possible NCFs for a node with $k=1$ input is 2 and with $k=3$ inputs is 64, and the total number of all possible combinations of NCFs for the minimal GRN of 5 genes is $2 \times 64^4 \approx 3.4 \times 10^7$ (Table 1). For the minimal GRN of 5 genes shown in Fig. 3C, we find the number of possible SGNs for the node with $k=1$ input to be 1, and for each of the 4 nodes with $k=3$ inputs to be 16. Thus, the total number of all possible combinations of SGNs for the minimal GRN of 5 genes is $1 \times 16^4 = 65536$ (Table 1). Note that while assigning SGNs to nodes which have some inputs with unknown nature of interaction (activation or repression), we allow both possibilities (positive or negative unateness) for such interactions with unknown signs in the SGN. In summary, as reported in Table 1, the number of possible Boolean functions for the minimal GRN of 5 genes decreases from $1.7 \times 10^{10}$ to $8.3 \times 10^{8}$ when restricting Boolean functions to CFs, to $3.4 \times 10^{7}$ when restricting to NCFs, and down to $6.5 \times 10^{4}$ when restricting to SGNs. Thus, restriction to SGNs significantly constrains the number of possible Boolean functions for the minimal GRN of 5 genes by reducing the functions to a fraction $3.8 \times 10^{-6}$ of the complete space (Table 1).

Based on experimental measurements, we can define three cell states (corresponding to the exocrine cells, the $\beta/\delta$ progenitor and the $\alpha$/PP progenitor) for the minimal GRN of 5 genes with each gene adopting binary values: 0 when the gene is not expressed in a cell lineage and 1 when the gene is expressed in a cell lineage, as shown in Fig. 3D. These three cell states or lineages (which are labelled attractor 1, 2, and 3 in Fig. 3D) must be stable attractor states of the BN model for pancreas development, and thus, represent a dynamical constraint on the network. Therefore, as a second constraint, we impose that the BN models for pancreas development based on topology of the minimal GRN of 5 genes must have at least three stable attractors with the gene expression patterns as defined in Fig. 3D. Here we further refine this condition into two cases: (i) BNs that have at least the three defined attractors; (ii) BNs that have exactly three defined attractors. In the first case, for *at least three attractors*, we find that the number of combinations of Boolean functions satisfying the dynamical constraint to be $\sim 1.0 \times 10^{6}$. Among these $\sim 1.0 \times 10^{6}$ combinations, there are 78,400 CFs, within which there are 9,216 NCFs, and 3,600 SGNs (Table 1). In the second case, for *exactly three attractors*, we find the number of combinations of Boolean functions satisfying the dynamical constraint to be 86,042. Among these 86,042 combinations, there are 3,741 CFs, 219 NCFs, and 109 SGNs (Table 1). Supplementary Table S2 lists the 109 combinations of SGNs for the minimal GRN of 5 genes satisfying the dynamical constraint of *exactly three attractors*.

Our results reported in Table 1 show that the condition of sign-compatible functions is the strictest constraint leading to significant decrease in the number of possible Boolean functions. We remark that our method of applying successive structural constraints (CFs, NCFs and SGNs) to shrink the space of possible Boolean functions can serve as a powerful tool to reconstruct large-scale BNs from experimental datasets. The logic relationship of various constraints for the nested subsets of Boolean functions is shown in the Venn diagram in Fig. 3E.

### 3.2 Imposing structural constraints enriches the biologically correct relative ordering of attractors

From experimental observations, it is known that among the cell types of pancreas studied here, the exocrine cells are the least stable while the $\beta, \delta$ progenitors are more stable than the $\alpha, \gamma$ progenitors [35]. In order to study the relative stability of cell states in pancreas differentiation, we demand that the corresponding BN model not only recapitulates the correct (observed) gene expression patterns for the three cell types, but also that it reproduces the correct *developmental ordering* of the three cell types, that is, of the three attractors. Thus, we impose one additional functional constraint for BN model determination: transition rates between the attractors should reflect the developmental ordering. Consideration of this additional functional constraint of developmental ordering within BN framework to build models raised a central question: when one imposes structural and dynamical constraints as in the last section to shrink the space of possible BNs, does it lead to the enrichment of correct relative ordering of cell attractors? Here we define the relative ordering of cell attractors in BN models of development based on MFPT which is used to measure the relative stability of attractor states (See Sections 2.5 and 2.6).

We employ an ensemble approach to determine whether imposition of increasing level of constraints on the set of possible BNs will increase the probability of obtaining the BNs with correct relative ordering. In other words, we decided to check if the set of BNs with correct relative ordering will be enriched when more stringent constraints are put on the ensemble of possible Boolean functions. To estimate the statistical significance, a null model was constructed to bootstrap the same number of Boolean functions as the ones defined by the structural constraints (CFs, NCFs and SGNs). For example, there are 9,216 NCFs among the set of 78,400 CFs satisfying the dynamical constraint of *at least three attractors*. In the null model, we use bootstrapping to repeatedly randomly choose 9,216 functions from the set of 78,400 CFs and calculated their ordering distribution based on MFPT to generate a Boxplot (Fig. 4). In Fig. 4, only the second panel represents the correct biological ordering (Stability of Attractor 1 < Stability of Attractor 3 < Stability of Attractor 2) for pancreas differentiation. From the Boxplot in Fig. 4, it is seen that when we impose increasing constraints on the type of Boolean functions in the order of CFs, NCFs and SGNs, the ratio of the BNs with correct relative ordering (represented by black dots), always outperforms the random sampling. Bootstrapping for other possible orderings show mixed results compared with random sampling. In fact, the opposite ordering (Stability of Attractor 2 < Stability of Attractor 3 < Stability of Attractor 1) is diluted as can be seen in panel 5 of Fig. 4. These observations further validate our choice of the *sign-compatible* functions, which not only significantly decrease the number of possible BNs, but also promote correct ordering of the relative stabilities of cell attractors in the GRN for pancreas development.

### 3.3 Bimodal distribution of Landscape between Flatness and Ruggedness

Besides building a BN model for the correct relative ordering of cell attractors, it is of general interest to characterize the level of effort required for a cell to transition from one state to another state, i.e., is the BN landscape of cell state transitions rather flat or rugged? A simple measure to evaluate the ruggedness of the BN landscape is the entropy $S = -\sum_i p_i^* \ln p_i^*$ associated with the steady state probability distribution $\mathbf{p}^*$ of the BN model with noise where elements $p_i^*$ give the probability for the system to be in state $s_i$. The larger the entropy $S$, the flatter is the landscape and *vice versa*. Since the entropy of the steady state probability distribution of the BN model varies with the noise level $\eta$, we computed the entropies of the 86,024 combinations of Boolean functions (producing exactly three attractors) for different noise levels $\eta \in [0.001, 0.009]$ in increments of 0.001, $\eta \in [0.01, 0.09]$ in increment 0.01, and $\eta \in [0.1, 0.7]$ in increments of 0.1, as plotted in Fig. 5A. We find that the variance of entropies of the Boolean function ensemble becomes smaller and the landscape becomes flatter with increasing noise. This implies that presence of large amplitude of noise in BN models diminishes the influences of constraints specified by GRN. Since all noise levels were generally found to have the same trend, we chose the noise level $\eta=0.01$ to plot the entropies of 86,024 steady state distributions corresponding to all combinations of Boolean functions with exactly three attractors in Fig. 5B (where different colors correspond to different relative ordering). From Fig. 5B, it is seen that the patterns of entropy across the combinations of Boolean functions with exactly three attractors is almost repeatable, but interestingly this is not due to degeneracy in steady state distributions. We have explicitly checked that each steady state distribution $\mathbf{p}^*$ is unique, with the overall trend that the entropy hops between two peaks when we sample all combinations of Boolean functions with exactly 3 attractors.

From Fig. 5B and 5C, we can see that the entropy distributions for different relative ordering schemes appear to be bimodal. Since the state space of the minimal GRN of 5 genes has 32 states, the absolute random case corresponds to the entropy $S = -\sum_i p_i^* \ln p_i^* = -\frac{1}{32} \ln \frac{1}{32} \cdot 32 \approx 3.46$. From Fig. 5B, it is seen that the maximum entropy for BNs of any relative ordering is below 1.8, suggesting that the selected networks were all in the ordered regime, far away from pure randomness. Note that smaller the value of entropy, greater is the level of constraints on a BN model. Thus, in Fig. 5B, the peak near zero reflects more rugged landscapes while the peak far away from zero denotes the flat landscapes. It is interesting to notice that even two dynamical systems with exactly the same attractors and the same relative ordering, can display a distinct overall landscape, either flat or rugged. Cell reprogramming is much more difficult in the rugged landscape than in the flat one because of the many intermediate attractors between the origin and the destination attractors. Thus, the above described scheme could be exploited for the analysis, and perhaps, rational protocol design of cell reprogramming. Specifically, one may be able to perturb the GRN such as to make the landscape flatter while maintaining its biological ordering of cell attractors.

We next tried to identify common features across the set of Boolean functions leading to rugged landscapes in the minimal GRN of 5 genes. Given any Boolean function, we define a quantity called *rule number* associated with the function that gives the number of Boolean rules separated by logical OR in the disjunctive normal form of the Boolean function. For example, the Boolean function $A$ AND NOT $B$ AND $C$ has rule number 1; the Boolean function (NOT $A$ AND $B$) OR ($B$ AND NOT $C$) has rule number 2. In Figs. 6A and 6B, we plot the distribution of rule numbers for Boolean functions at each node of the minimal GRN of 5 genes for both the flat and the rugged landscapes. The main difference between the two types of landscapes we found was that the Boolean functions of the rugged landscape had, on average, higher rule numbers than those of the flat ones. One possible explanation for this finding is as follows: In case of dynamical systems with the same set of attractors, more rule numbers in Boolean functions leads to a more rugged landscape, which in turn renders more difficult the cell state transitions.

## 4. Conclusions

In the post-genome era, understanding of biological processes has shifted from a notion of causation focused on few genes or linear regulatory pathways to one that takes into account the high-dimensional dynamics of complex nonlinear systems, prosaically represented by GRNs. Many higher-level cell functions, such as cell differentiation, cell cycle, immune responses or neuronal activities can only be explained by the dynamics of complex biological networks. In this paper, we built on the old elementary concept of the cell attractor that is now gaining attention and showed that Waddington's epigenetic landscape has a formal basis even in the BN framework. Porting the landscape concept to the realm of discrete networks allowed us to use computational tools to study the landscape as additional information derived from the GRN. We defined the relative stability of network states on the BN landscape, thus providing a formal and quantifiable basis to the elevation in the *landscape for Boolean network*. We proposed methods to enforce the relative ordering of attractor states in BN models as a novel requirement for constraining large ensembles of BNs to the biologically relevant class. We showed in our example that even with incomplete information about network structures, the use of BN can capture the essential dynamics of cell fate change and permit the estimation of relative stability and transition barriers between the cell attractors. This mathematical framework has the potential to assist the rational design of perturbation protocols for directing cell reprogramming in order to generate the desired cell lineages in regenerative medicine. As the knowledge of the structure of GRNs governing the development of various tissues increases in the next decade, the utilization of such network information for therapeutic reprogramming may benefit from the concepts developed here.

**Acknowledgements**

Research reported in this publication was supported by the Center for Systems Biology/2P50GM076547 of the National Institutes of Health under award R01GM987654. This research was also supported by the National Science Foundation Grant PHY11-25915. The content is solely the responsibility of the authors and does not necessarily represent the official views of the funding agencies.

**Figures**

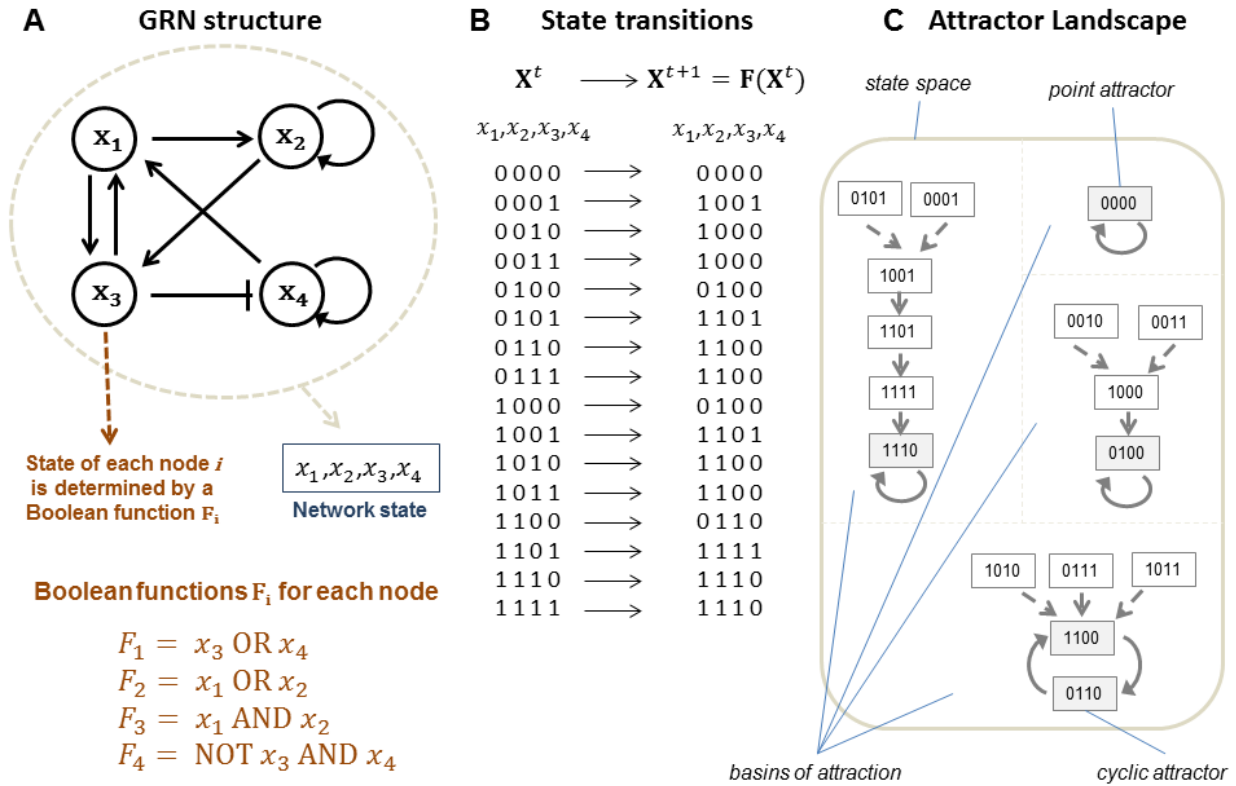

**Figure 1: Example of a BN model and its attractor landscape.** (a) Structure of the BN model with four genes along with their interactions. Different types of edges are used to distinguish between the different types of gene interactions (activation or inhibition). Also shown are the set of Boolean functions governing the output state of each node in the network. (b) The attractor structure for the BN model can be determined by explicitly evaluating state transitions for the 16 possible states of the network. (c) Attractor landscape of the BN model consisting of 4 attractors and associated basins of attraction within state space. Arrows depict transitions between two states in the state space which are governed by the set of Boolean functions at each node in the network. Stable states (fixed-point attractors) transition to themselves while cyclic attractors oscillate within a subset of states following a certain time sequence.

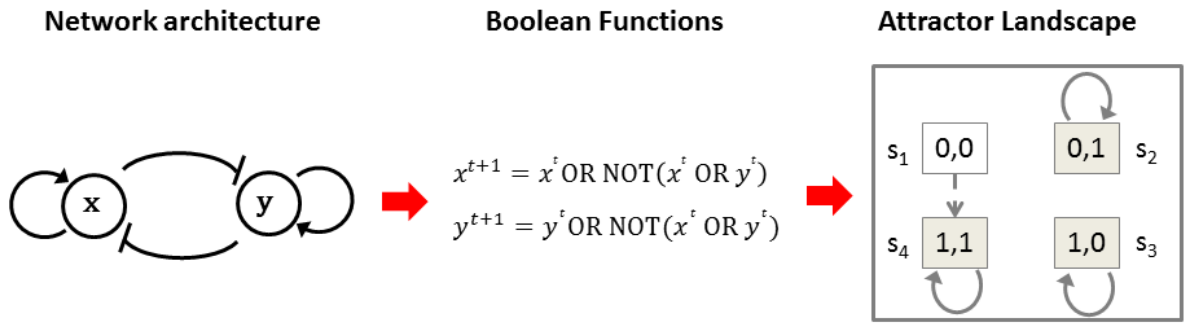

**Figure 2: BN model for a toggle switch of two genes.** For the 2-gene toggle switch, the figure shows the network structure with cross-inhibition and self-activation, the Boolean functions for the two genes, and the attractor landscape for the BN model.

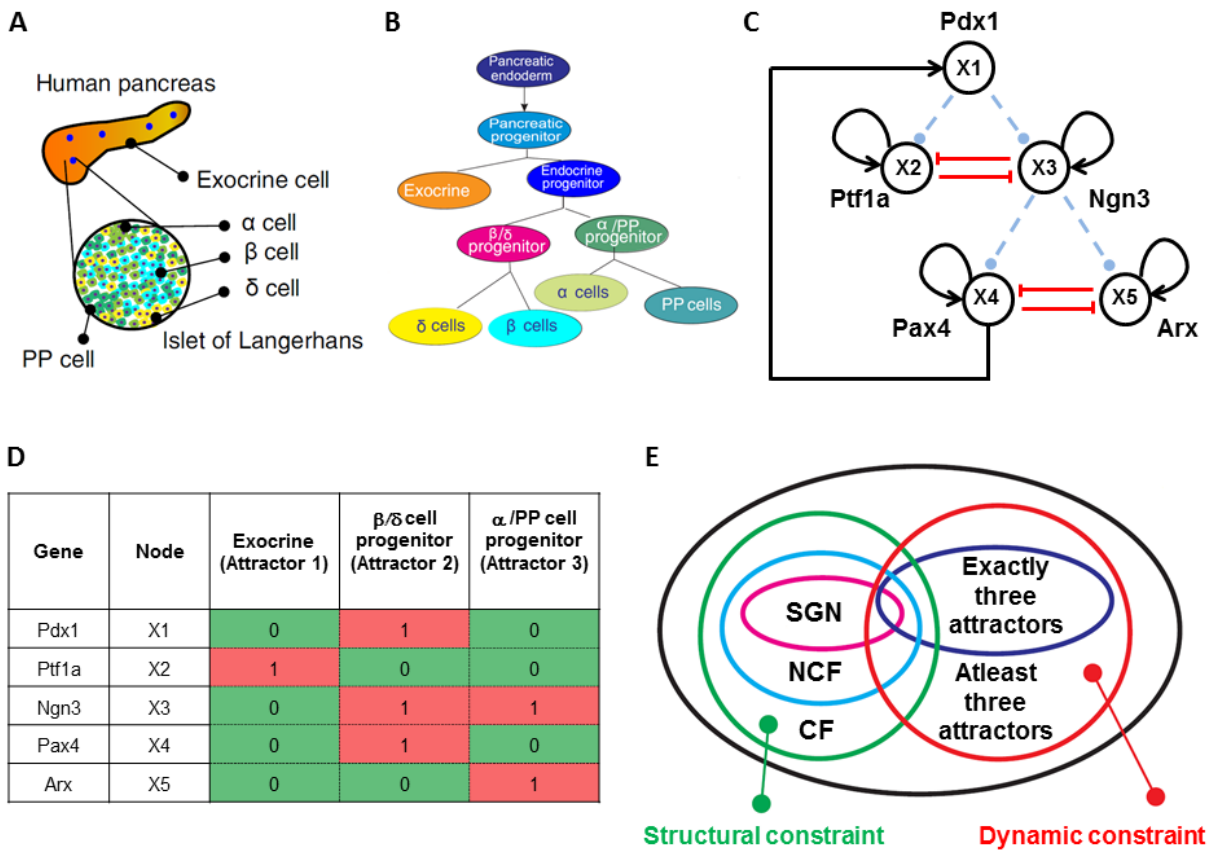

**Figure 3: BN model of the minimal GRN for Pancreas cell differentiation. (a)** The anatomical structure of the pancreas which consists of the exocrine cells and the islets of Langerhans with the endocrine cells. $\beta$ cells in the islets secrete insulin to regulate the blood sugar level. **(b)** Developmental tree of pancreas cells. Pancreas progenitor cells first differentiate

into the exocrine cells and endocrine progenitors which can further differentiate into β/δ progenitors and α/PP progenitors. β/δ progenitors differentiate into δ cells and β cells. α/PP progenitors differentiate into α cells and PP cells. **(c)** Structure of the minimal GRN model recapitulating the pancreas developmental process for the three main lineages (exocrine, β/δ and α/PP) considered here. The network mainly consists of two layers of two mutually-inhibitory fate genes. The first cross-inhibition between *Ptf1a* and *Ngn3* determines the switch between the exocrine cells and endocrine progenitors. The second cross-inhibition between *Pax4* and *Arx* determines the switch between the β/δ progenitors and α/PP progenitors. Here, black edges terminating with arrow heads represent known activating interactions, red edges terminating with bars represent known inhibitory interactions, and blue dashed edges terminating with circles represent interactions with unknown signs. **(d)** The gene expression patterns corresponding to the three cell attractors of the minimal GRN for pancreas differentiation. Attractor 1 corresponds to exocrine cells, 2 to β/δ progenitors and 3 to α/PP progenitors. **(e)** Venn diagram shows that imposition of successive structural constraints (CFs, NCFs and SGNs) and dynamical constraint of three stable attractor states corresponding to cell lineages limits the space of possible Boolean functions for the minimal GRN of 5 genes.

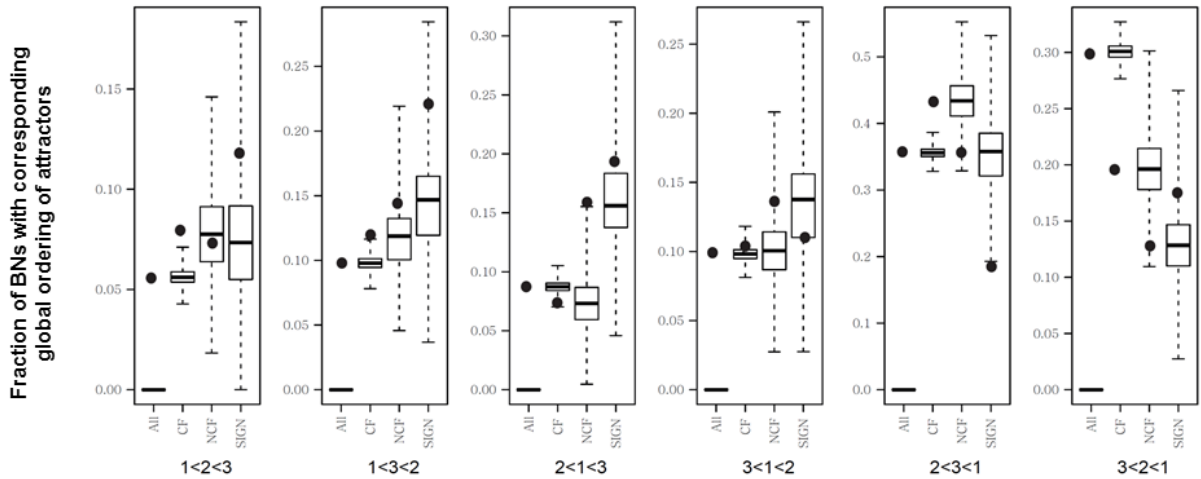

**Figure 4: Distribution of BN models satisfying increasing structural constraints and the dynamical constraint of exactly three defined attractors across different possible relative stability orderings of cell attractors.** This Boxplot has six panels with each corresponding to a different relative ordering of cell attractors. For example, the panel labeled 1<2<3 corresponds to the relative ordering where the relative stability of attractor 1 is less than that of attractor 2 which is less than that of attractor 3. Attractors 1, 2 and 3 correspond to exocrine cells, the β/δ progenitor and the α/PP progenitor, respectively. The second panel (1<3<2) corresponds to the *correct* (experimentally observed) ordering of cell states. In each panel, the black dots represent

the fraction of BN models having the corresponding relative ordering of attractors in addition to satisfying the dynamical constraint of exactly three defined attractors and one of the following structural constraints w.r.t. Boolean functions: ALL – all Boolean functions; CF – canalyzing functions; NCF – nested canalyzing functions; and SGN – sign-compatible functions. The distributions in different panels of the Boxplot give the expected value from the Null model.

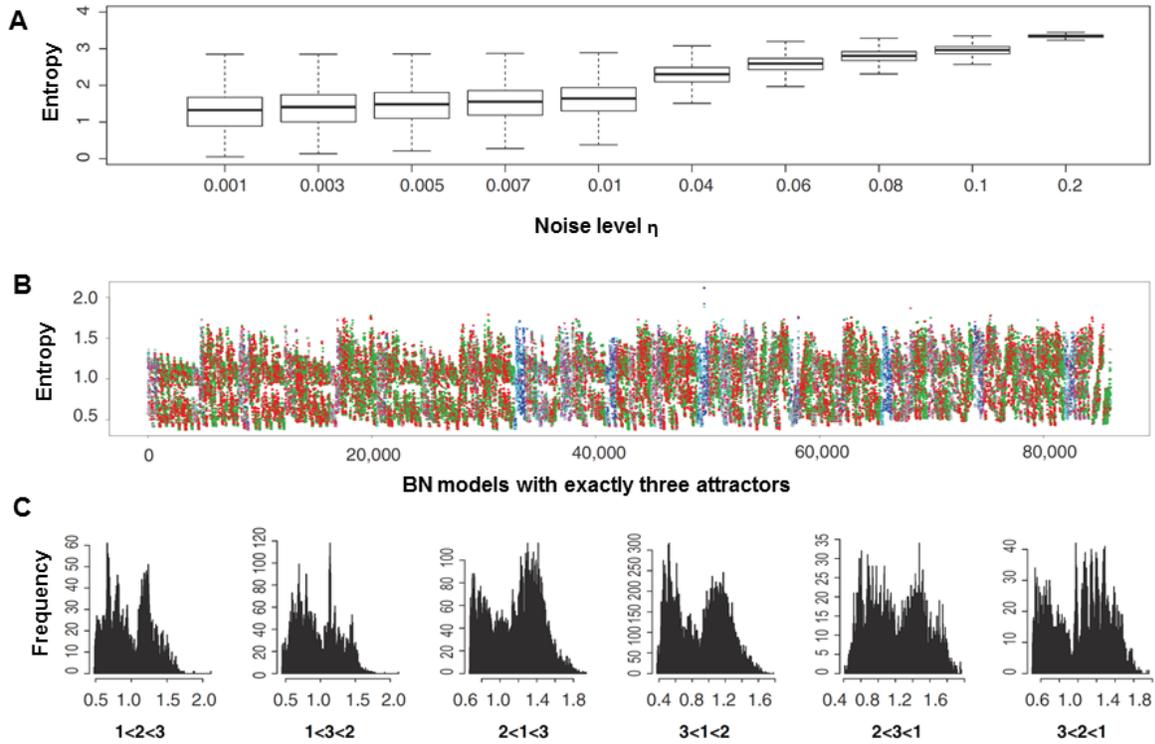

**Figure 5: Entropy associated with the steady state probability distribution of BN models satisfying structural and dynamical constraints. (a)** Entropy distribution for 86,024 BN models of pancreas development satisfying the dynamical constraint of exactly three attractors for various levels of noise $\eta$. Results are shown for ten different noise levels $\eta$ in the range from 0.001 to 0.2. With increase in noise levels $\eta$, the nature of distribution changes from broad with values in the range [0,3] to that approaching an absolute chaos with entropy $S = -\sum_i p_i^* \ln p_i^* = -\frac{1}{32} \ln \frac{1}{32} \cdot 32 \approx 3.46$. **(b)** Entropy distribution for 86,024 BN models of pancreas development satisfying the dynamical constraint of exactly three attractors with noise level $\eta = 0.01$ and different relative ordering of attractor states. In this figure, the color scheme for different relative ordering of attractors is as follows: *Blue* - 1<2<3; *Cyan* - 1<3<2; *Red* - 2<1<3; *Green* - 3<1<2; *Magenta* - 2<3<1; *Grey* - 3<2<1. **(c)** Histogram plot of entropy distribution for 86,024 BN models of pancreas development satisfying the dynamical constraint of exactly three attractors

with noise level $\eta = 0.01$ across different relative ordering of attractor states. It is seen that distributions are distinctively bi-modal.

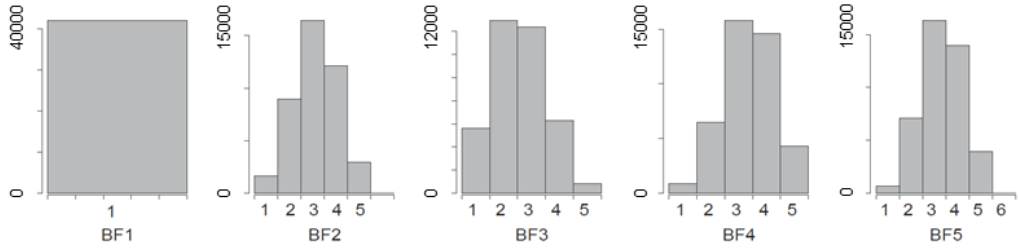

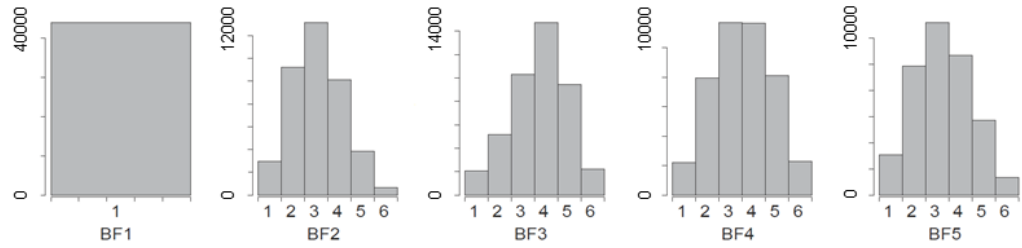

**Figure 6: Distribution of rule numbers for Boolean functions at each node of the minimal GRN of 5 genes for flat and rugged landscapes. (a)** Distribution for flat BN landscapes. **(b)** Distribution for rugged BN landscapes. In this figure, $BF_i$ refers to Boolean function for gene $x_i$ in the minimal GRN of 5 genes.

# Tables

**Table 1: Number of possible combinations of Boolean functions satisfying structural and dynamical constraints associated with BN model of pancreas development.**

| Type of Boolean functions | Fixed network structure but without any dynamical constraint | Fixed network structure and dynamical constraint of at least three known attractors | Fixed network structure and dynamical constraint of exactly three known attractors |
|---|---|---|---|
| All Boolean functions (ALL) | $1.71799 \times 10^{10}$ | 1,048,576 | 86,024 |
| Canalyzing functions (CF) | $8.2944 \times 10^{8}$ | 78,400 | 3,741 |
| Nested canalyzing functions (NCF) | $3.35544 \times 10^{7}$ | 9,216 | 219 |
| Sign-compatible functions (SGN) | 65,536 | 3,600 | 109 |

# Supplementary Material

**Supplementary Table S1:** Literature evidence in support of interactions between genes in the minimal GRN for pancreas development.

**Supplementary Table S2:** List of 109 combinations of sign-compatible functions (SGNs) for the 5 gene minimal GRN for pancreas development satisfying the dynamical constraint of exactly three attractors.